\documentclass[journal]{IEEEtran}
\usepackage{algorithm}
\usepackage{algorithmicx}
\usepackage{algpseudocode}
\usepackage{amsmath}
\usepackage{mathrsfs}
\usepackage{amsfonts}
\usepackage[numbers,sort&compress]{natbib}
\usepackage{amssymb}
\usepackage{bm}
\usepackage{enumerate}
\usepackage[pdftex]{graphicx}
\usepackage{lineno}
\emergencystretch=\maxdimen
\ifCLASSINFOpdf
\else
\fi

\begin{document}
%
% paper title
% Titles are generally capitalized except for words such as a, an, and, as,
% at, but, by, for, in, nor, of, on, or, the, to and up, which are usually
% not capitalized unless they are the first or last word of the title.
% Linebreaks \\ can be used within to get better formatting as desired.
% Do not put math or special symbols in the title.
\title{Multiple Description Convolutional Neural\\ Networks for Image Compression}
%
%
% author names and IEEE memberships
% note positions of commas and nonbreaking spaces ( ~ ) LaTeX will not break
% a structure at a ~ so this keeps an author's name from being broken across
% two lines.
% use \thanks{} to gain access to the first footnote area
% a separate \thanks must be used for each paragraph as LaTeX2e's \thanks
% was not built to handle multiple paragraphs
%

\author{Lijun~Zhao,
        Huihui~Bai,~\IEEEmembership{Member,~IEEE,}
        Anhong~Wang,~\IEEEmembership{Member,~IEEE,}
        and Yao~Zhao,~\IEEEmembership{Senior~Member,~IEEE}
        %and~Ce~Zhu,~\IEEEmembership{Fellow,~IEEE}

\thanks{L.  Zhao, H. Bai, Y. Zhao are with the Beijing Key Laboratory of Advanced Information Science and Network Technology, Institute Information Science, Beijing Jiaotong University, Beijing, 100044, P. R. China, e-mail: {15112084, hhbai, yzhao}@bjtu.edu.cn.}
\thanks{A. Wang is with Institute of Digital Media \& Communication, Taiyuan University of Science and Technology, Taiyuan, 030024, P. R. China, e-mail: wah\_ty@163.com}
%\thanks{C. Zhu is with the School of Electronic Engineering, University of Electronic Science and Technology of China, Chengdu 611731, China, and also with the Center for Robotics, University of Electronic Science and Technology of China, Chengdu 611731, China, e-mail: eczhu@uestc.edu.cn}
%\thanks{Manuscript received X XX, 20XX; revised X XX, 20XX.}
}
% note the % following the last \IEEEmembership and also \thanks -
% these prevent an unwanted space from occurring between the last author name
% and the end of the author line. i.e., if you had this:
%
% \author{....lastname \thanks{...} \thanks{...} }
%                     ^------------^------------^----Do not want these spaces!
%
% a space would be appended to the last name and could cause every name on that
% line to be shifted left slightly. This is one of those "LaTeX things". For
% instance, "\textbf{A} \textbf{B}" will typeset as "A B" not "AB". To get
% "AB" then you have to do: "\textbf{A}\textbf{B}"
% \thanks is no different in this regard, so shield the last } of each \thanks
% that ends a line with a % and do not let a space in before the next \thanks.
% Spaces after \IEEEmembership other than the last one are OK (and needed) as
% you are supposed to have spaces between the names. For what it is worth,
% this is a minor point as most people would not even notice if the said evil
% space somehow managed to creep in.

% The paper headers
\markboth{Journal of \LaTeX\ Class Files}%,~Vol.~xx, No.~x, x~20xx
{Shell \MakeLowercase{\textit{et al.}}: Bare Demo of IEEEtran.cls for IEEE Journals}
% The only time the second header will appear is for the odd numbered pages
% after the title page when using the twoside option.
%
% *** Note that you probably will NOT want to include the author's ***
% *** name in the headers of peer review papers.                   ***
% You can use \ifCLASSOPTIONpeerreview for conditional compilation here if
% you desire.

% If you want to put a publisher's ID mark on the page you can do it like
% this:
%\IEEEpubid{0000--0000/00\$00.00~\copyright~2015 IEEE}
% Remember, if you use this you must call \IEEEpubidadjcol in the second
% column for its text to clear the IEEEpubid mark.

% use for special paper notices
%\IEEEspecialpapernotice{(Invited Paper)}

% make the title area
\maketitle

% As a general rule, do not put math, special symbols or citations
% in the abstract or keywords.
\begin{abstract}
Multiple description coding (MDC) is able to stably transmit the signal in the un-reliable and non-prioritized networks, which has been broadly studied for several decades. However, the traditional MDC doesn't well leverage image's context features to generate multiple descriptions. In this paper, we propose a novel standard-compliant convolutional neural network-based MDC framework in term of image's context features. Firstly, multiple description generator network (MDGN) is designed to produce appearance-similar yet  feature-different multiple descriptions automatically according to image's content, which are compressed by standard codec. Secondly, we present multiple description reconstruction network (MDRN) including side reconstruction networks (SRN) and central reconstruction network (CRN). When any one of two lossy descriptions is received at the decoder, SRN network is used to improve the quality of this decoded lossy description by removing the compression artifact and up-sampling simultaneously. Meanwhile, we utilize CRN network with two decoded descriptions as inputs for better reconstruction, if both of lossy descriptions are available. Thirdly, multiple description virtual codec network (MDVCN) is proposed to bridge the gap between MDGN network and MDRN network in order to train an end-to-end MDC framework. Here, two learning algorithms are provided to train our whole framework. In addition to structural similarity loss function, the produced descriptions are used as opposing labels with multiple description distance loss function to regularize the training of MDGN network. These losses guarantee that the generated description images are structurally similar yet finely diverse. Experimental results show a great deal of objective and subjective quality measurements to validate the efficiency of the proposed method.
\end{abstract}

% Note that keywords are not normally used for peerreview papers.

\IEEEpeerreviewmaketitle
\section{Introduction}
% The very first letter is a 2 line initial drop letter followed
% by the rest of the first word in caps.
%
% form to use if the first word consists of a single letter:
% \IEEEPARstart{A}{demo} file is ....
%
% form to use if you need the single drop letter followed by
% normal text (unknown if ever used by the IEEE):
% \IEEEPARstart{A}{}demo file is ....
%
% Some journals put the first two words in caps:
% \IEEEPARstart{T}{his demo} file is ....
%
% Here we have the typical use of a "T" for an initial drop letter
% and "HIS" in caps to complete the first word.
\IEEEPARstart{L} {arge} amounts of attentions have been paid to various techniques of Internet service and multimedia signal transmission for many years, which not only provide us a convenient manner of communication but also give us many choices for our life style. Meanwhile, the bandwidth of Internet has been accelerated and more stable transmission service is guaranteed by these developments. But there are still some risks of transmission failures, when the Internet congestion occurs in the overloaded case or signal packets are conveyed in the unpredictable yet unreliable channels \cite{m43,m44}. Multiple description coding has been studied as a promising technique of source coding to relieve these problems by decomposing the signal into multiple redundant subsets, which are transmitted in different channels. Thus, a degraded but acceptable signals reconstruction can be produced after decoding, even though only one description is received at the clients. If more descriptions are available for users, better quality of signal reconstruction can be achieved. Multiple description coding has been widely explored in the field of image and video coding \cite{m1, m33, lml1, m9, m11, lml2, m5, m18, m19, m32, m10, m13, m14, m6, m12, m8, m7, m16, m20}.

As one of the main techniques in multiple description image coding, multiple description scalar quantization could overcome impairments of transmission channel \cite{m9}. For example, in \cite{m11}, multiple description scalar quantizers have been combined with efficient wavelet coders to generate independent multiple packets for error resilience. In \cite{lml1}, two-stage multiple description scalar quantization is presented to create central and side decoders, whose distortions are closer to the rate-distortion bound of multiple description coding under the condition of the high-resolution assumption. To cope with the L-description problem\cite{m5}, two novel coding schemes are proposed, when the symmetric rates and symmetric distortion is constrained. In \cite{m1}, a new achievable rate-distortion region with combinatorial message sharing is presented by introducing shared codebooks and the refinement codebook to generate L-channel multiple descriptions.

Compared with multiple description scalar quantization, lattice vector quantization characterizes in good symmetric structure of lattices and avoiding complex nearest neighbor searching. In \cite{m18}, the main problem of designing lattice vector quantizer is formulated as a labeling problem for two-channel multiple description. In \cite{m19}, non-lattice codebook with symmetries of the coarse lattice is used to get objective quality gains for multiple description coding but without a great increase of complexity. In \cite{m32}, multiple description lattice vector quantization is operated in an optimized way in terms of appropriate construction of wavelet coefficient vectors, choosing sub-lattice index values and different subbands quantization step on the wavelet domain. In \cite{lml2}, the index assignment of multiple description lattice vector quantization is designed to be translated into a transportation problem and greedy algorithm as well as general algorithms is developed to pursue optimality of the index assignment.

Except multiple descriptions directly produced by quantization, there are many alternative strategies for multiple description coding. To generate two descriptions in transform based coding framework, correlation between pairs of transform coefficients is introduced by a pairwise correlating transform \cite{m10}. This correlation facilities to reduce the distortion when only a single description is received. Later, both domain-based multiple description coding and forward error correction are used for concatenated multiple description coding of frame-rate scalable video \cite{m13}. Meanwhile, both prioritized discrete cosine transform in video compression and multiple description codes based on forward error correction are combined together to provide a wireless channels video transmission scheme \cite{m14}.

From literatures \cite{m13,m14}, it can be observed that multiple description video coding using forward error correction has been widely explored. There are several other kinds of multiple description video coding. In \cite{m6}, a video is coded into multiple independently streams so that each stream has its own prediction and dependent state to defeat against bit error or packet loss.  In multiple description motion coding algorithm, motion vector is encoded into two descriptions, which are transmitted over distinct channels to the decoder so that motion vector field is robust against transmission errors \cite{m12}. In the scalable wavelet video codec, each packet is encoded with a separate channel code, so that the integrity of the packets is protected and it allows to detect packet-decoding failures cases, after breaking wavelet transformation into several spatial-temporal tree blocks \cite{m8}. In \cite{m7}, two architectures of multiple description video coding are built up based on motion compensation prediction loop and a poly-phase down-sampling technique is chosen to generate multiple descriptions and introduce cross redundancy among the descriptions.

Although the aforementioned approaches can well alleviate the congestion of Internet and satisfy the demanding of real-time application, these approaches are not compatible to standard codec, such as JPEG, and JPEG2000. To resolve this problem, some previous works have provided some feasible solutions, such as \cite{m16, m20, m33, lml2}. In \cite{m20}, through grouping the codeblock to generate two balanced set, these two set are compressed by JPEG-2000 with two different quantization parameter to get four subsets, which are interlacedly merged together to create two descriptions. In \cite{m16}, the rate-allocation strategy embedded in the JPEG2000 encoder is introduced for the rate-distortion optimization of multiple descriptions of images, in which single description decoding is able to compatible with JPEG2000 Part 1 decoder. In view of human eyes' always sensitivity to the changes above just noticeable difference (JND) threshold, only the significant visual information, which contributes to the JND tolerance, is encoded as the redundant information during H.264/AVC based multiple description video coding \cite{m33}. In \cite{lml2}, frame-level rate-distortion optimized description generation scheme takes account of temporal coding dependency to minimize the end-to-end distortion, which is built on standard H.264/AVC.

Because the proposed approach is high related about the issue of compression artifact removal \cite{m21,m22,m24,m23,m35,m27,m28,m29,m36}, we will next review several state-of-the-art works about compression artifact removal. In \cite{m21}, pointwise shape-adaptive discrete cosine transform is leveraged for both denoising and deblocking after image compression. In \cite{m22}, dictionary learning is introduced to reducing JPEG-compressed artifacts in view of image's sparse and redundant representations. In \cite{m24}, collaborative filtering is designed to uncover the finest details and maintain each individual block's unique features in the sparse 3-D transform-domain, which is not restrict to the denoising of compressed image, so this approach is a general denoising method. Lately, the deblocking problem is formulated as an optimization problem, where non-convex low-rank model constrained is considered to reduce blocking artifacts \cite{m23}. Meanwhile, the popular techniques of convolutional neural network and generative adversarial have been tried to remove artifacts \cite{m27, m28, m29}.

Following the work of \cite{m7}, we form multiple description coding baselines with a poly-phase down-sampling technique to generate multiple descriptions by combining state-of-the-art artifact removal technique with super-resolution based on very deep convolutional neural network. Specifically, the input image is down-sampled with a poly-phase down-sampling technique along the main diagonal for each $2\times 2$ non-overlapped window to form two descriptions for coding with standard codec. After decoding, several state-of-the-art artifact removal techniques, such as \cite{m21,m22,m23,m24} are used to enhance image quality, which is followed by super-resolution to restore image from low-resolution to high-resolution with very deep convolutional neural network, such as novel super-resolution methods of \cite{m25} and \cite{m45}. The combinatorial methods with artifact removal \cite{m21,m22,m23,m24} and super-resolution \cite{m25} are respectively referred to as multiple description coding baselines1-4, namely "MDB1a", "MDB2a", "MDB3a", "MDB4a". In this similar way, when artifact removal methods of \cite{m21,m22,m23,m24} are combined together with \cite{m45}, they are respectively denoted as "MDB1b", "MDB2b", "MDB3b", "MDB4b".

\begin{figure*}[ht]
\centering
\includegraphics[width=6.3in]{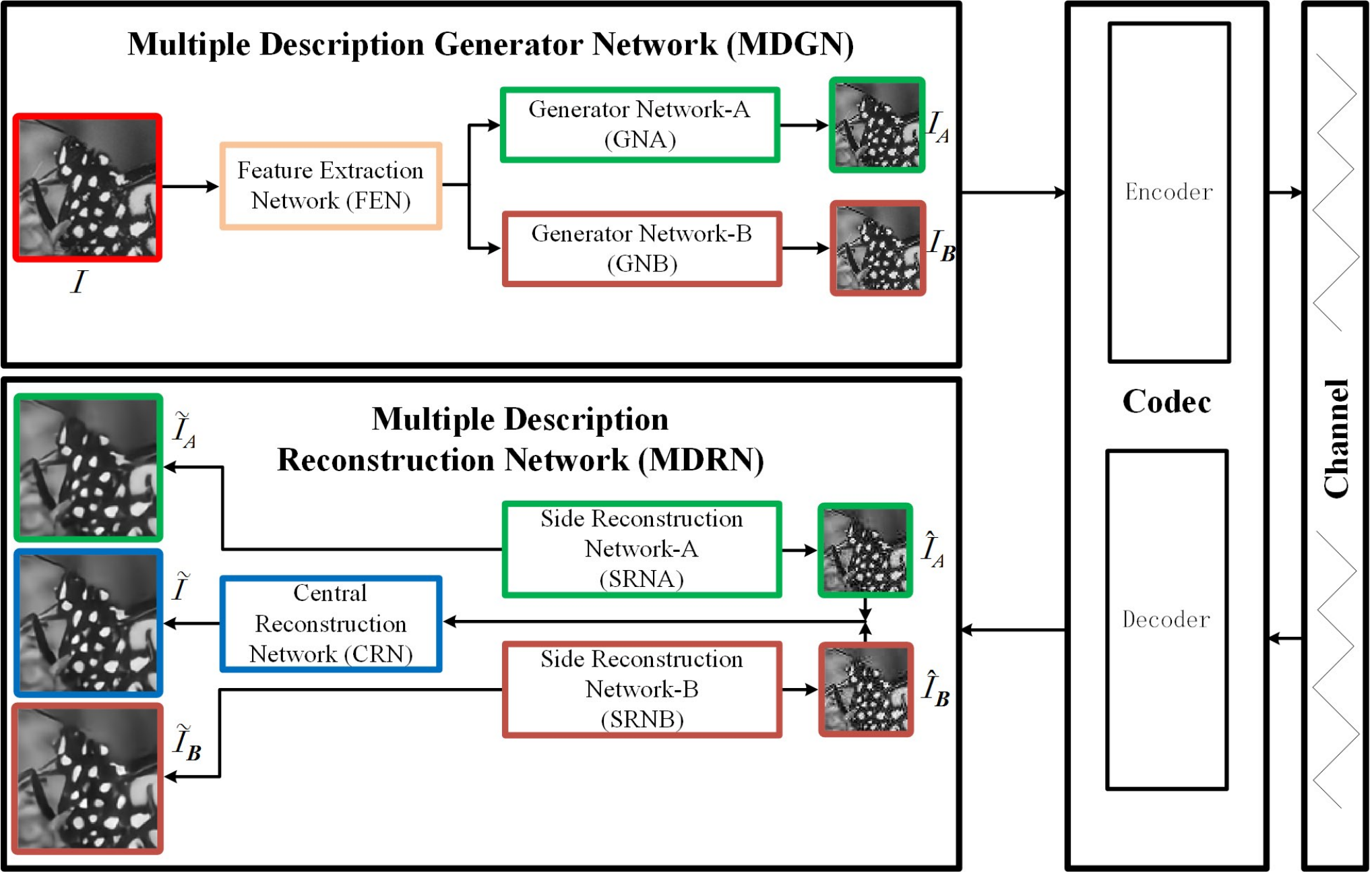}
\caption{The framework of multiple description coding based on deep convolutional neural networks}
\label{Fig1}
\end{figure*}

In this paper, we introduce a novel standard-compatible multiple description coding framework, in which multiple descriptions are produced by deep convolutional neural network. Our contributions are listed as follows:
\begin{itemize}
\item Multiple description generator network (MDGN) is introduced to adaptively generate multiple descriptions according to image's content, which are compressed by standard codec to reduce transmission bits.
\item We present multiple description reconstruction network (MDRN), which consists of side reconstruction networks (SRN) and central reconstruction network (CRN). When either one of two compressed description is received at the decoder, side reconstruction network-A network (SRNA) or side reconstruction network-B (SRNB) is used to reconstruct the lossy description and enlarge this description simultaneously by removing compression artifact and up-sampling. Meanwhile, we utilize CRN network with two received descriptions as inputs to achieve high-quality image reconstruction, if all the multiple description images are available.
\item We train the aforementioned two neural networks: MDGN network and MDRN network together by learning multiple description virtual codec network (MDVCN).  It means that the learned MDVCN network is leveraged to further supervise the MDGN network's training. Besides, we provide two kinds of learning algorithms for training our convolutional neural networks.
\item Distance loss for MDGN network is introduced as well as structural similarity loss to guarantee that the generated description images are structurally similar yet finely different.
\end{itemize}
The rest of this paper is given as follow. We first introduce the proposed methodology in Section 2. After that, we conduct a series of the experimental results to validate the efficiency in the Section 3. At last, we give a conclusion in the Section 4.

\section{The methodology}
In this paper, multiple description coding framework based on deep convolutional neural network is introduced to efficiently compress images, when facing an unpredictable and non-prioritized channel. Our main works are put on how to generate multiple descriptions in terms of redundancy between each description and description's diversity for better central reconstruction. Meanwhile, we design the neural network for description's generation and reconstruction and introduce how to train our convolutional neural networks together used in the proposed method. To the best of our knowledge, this is the first work using convolutional neural network for multiple description coding.

\subsection{Framework}
Our multiple description coding framework has three components: MDGN network, standard codec of JPEG, MDRN network, as depicted in the Fig. \ref{Fig1}. The MDGN network $g(\bm{I},\omega)$ is responsible to generate diverse descriptions $\bm{I_A}$ and $\bm{I_B}$ from the ground-truth image $\bm{I}$ with size of $M\times N$. Here, $\omega$ is the parameter set of MDGN network and other networks' parameter set can be defined in this similar way. Due to the widely usage of standard codec, such as JPEG, the standard-compatible coding framework becomes significant for practical applications. Thus, we use the JPEG codec to compress these descriptions so that image redundancy can be further reduced to get the lossy descriptions $\bm{\hat{I}_A}$ and $\bm{\hat{I}_B}$. The JPEG compressions of $\bm{I_A}$ and $\bm{I_B}$ are respectively represented as $\bm{\hat{I}_A}=c(\bm{I_A},\phi)$, and $\bm{\hat{I}_B}=c(\bm{I_B},\phi)$, where $c(\cdot)$ is the compression function of codec. The compressed description streams are separately transmitted over different channels. However, image compression with standard codec often incurs coding artifacts. Thus, MDRN network, denoted as reconstruction function $R(\cdot)$, is leveraged to remove these artifacts for image enhancement and enlarge the lossy description so that the final reconstruction image is guaranteed to have the same size with the ground-truth image $\bm{I}$. Finally, the receiver can still decode the received packet to get a description for acceptable quality reconstruction $\bm{\tilde{I}_A}$ with SRNA network or $\bm{\tilde{I}_B}$ with SRNB network, even though any one description is missing, as displayed in the Fig. \ref{Fig1}. If both descriptions are received, high quality reconstruction $\bm{\tilde{I}}$ can be built by CRN network.

As we all know, it's not easy to jointly train the MDGN network and MDRN network, because the quantization function in the codec of lossy compression is non-differentiable. Thus, the reconstruction error from the MDRN network can't be directly back-propagated to the MDGN network. Following our previous work \cite{m30}, we learn the MDVCN network to imitate the two consecutive procedures of codec's compression and description's reconstruction with MDRN network. As a result, we can train our whole framework in an end-to-end fashion.

\subsection{Objective function}
The objective function for our multiple description coding framework is written as follows:
\begin{align}%\notag\\
&\mathop{\arg\min}_{\omega, \alpha, \theta} L_{MDGN}(\bm{I_A},\bm{I_B},\bm{I},\omega)\notag\\
&+L_{MDRN}(\bm{I_A},\bm{I_B},\bm{I},\alpha)+L_{MDVCN}(\bm{I_A},\bm{I_B},\theta),
\end{align}
\begin{align}%\notag\\
\alpha=[\alpha_1,\alpha_2,\alpha_3],\notag\\
\theta=[\theta_1,\theta_2,\theta_3],\notag\\
[\bm{I_A},\bm{I_B}]= g(\bm{I},\omega),
\end{align}
%\begin{equation}
%\begin{split}
%\mathop{\arg\min}_{\omega, \alpha, \theta} L_{MDGN}(\bm{I_A},\bm{I_B},\bm{I},\omega)+L_{MDRN}(\bm{I_A},\bm{I_B},\bm{I},\alpha)\\
%+L_{MDVCN}(\bm{I_A},\bm{I_B},\theta),\\
%\alpha=[\alpha_1,\alpha_2,\alpha_3],\\
%\theta=[\theta_1,\theta_2,\theta_3],\\
%[\bm{I_A},\bm{I_B}]= g(\bm{I},\omega),
%\end{split}
%\label{eqn::all}
%\end{equation}
where three losses for training are respectively the loss of MDGN network, the loss of MDRN network, and MDVCN network's loss.
\begin{align}%\notag\\
&L_{MDGN}(\bm{I_A},\bm{I_B},\bm{I},\omega)=L_{SSIM} (u(\bm{I_A}),\bm{I})\notag\\
&+L_{SSIM}(u(\bm{I_B}),\bm{I})+\beta∙L_{dis}(\bm{I_A},\bm{I_B}),\notag\\
&\beta=clip(0.2/QF,\kappa1,\kappa2)
\end{align}

The loss of $L_{MDGN}(\bm{I_A},\bm{I_B},\bm{I},\omega)$ is used to supervise the learning of the parameters $\omega$ of the MDGN network in Eq. (3), where $u(\cdot)$ is the linear up-sampling function and $\beta$ balances the contributions between descriptor's SSIM loss \cite{m41} and distance loss, which are in effect contradictory to a certain extent. In addition, $QF$ is the quality factor for JPEG compression and $clip(\cdot)$ is the clip function to restrict value between $\kappa1$ and $\kappa2$ (e.g., $\kappa1=5*10^{-3}$ and $\kappa1=5*10^{-2}$). Hence, the parameter of $\beta$ plays a significant role on generating valid multiple descriptions. Note that the better quality is encoded, when the larger $QF$ is set for JPEG.
%When QF is set to be very small, there are great distortions for compressed descriptions. At this condition, the MDGN network's training for generation of description will be ruined, if the assumption is that $\beta$ is very small, that is to say, the most of contribution is assigned to distance loss for the supervision of MDGN network, as compared to other losses. However, large distortion of compressed descriptions will lead to small contribution of distance loss, which goes against our assumption distance loss can be large. Thus, large value of $\beta$ should be assigned for the balance between descriptor's SSIM loss and distance loss, when small value of QF is chosen.

On one hand, we hope that the two produced descriptions structurally similar to the input image so that the decoded descriptions can be watched directly for receiver, even without the processing of MDRN network. Consequently, SSIM loss function is used to supervise each description's learning. For example, the SSIM for description $\bm{I_A}$ is defined as follows:
\begin{equation}
\begin{split}
L_{SSIM}(u(\bm{I_A}),\bm{I})=-\frac{1}{M \cdot N} \sum_{i} L_{SSIM}(u(\bm{I_A})_i,\bm{I}_i),
\end{split}
\label{eqn::SSIMLOSS}
\end{equation}
\begin{align}%\notag\\
&L_{SSIM}(u(\bm{I_A})_i,\bm{I}_i)=\notag\\
&\frac{(2\mu_{u(\bm{I_A})_i}\cdot \mu_{\bm{I}_i}+c1)(2\sigma_{u(\bm{I_A})_i \bm{I}_i}+c2)}{(\mu^2_{u(\bm{I_A})_i}+\mu^2_{\bm{I}_i}+c1)(\sigma^2_{u(\bm{I_A})_i}+\sigma^2_{\bm{I}_i}+c2)},
\end{align}
where $\mu_{u(\bm{I_A})_i}$ and $\sigma^2_{u(\bm{I_A})_i}$ respectively denote the mean value and the variance of the neighborhood window centered by pixel $i$ in the image $u(\bm{I_A})$. Similarly, $\mu_{\bm{I}_i}$ as well as $\sigma^2_{\bm{I}_i}$ is denoted in this way. $\sigma_{u(\bm{I_A})_i \bm{I}_i}$ is the covariance between neighbourhood windows centered by pixel $i$ in the image $u(\bm{I_A})$ and in the image $\bm{I}$. Meanwhile, $c1$ and $c2$ are two constant values (e.g., $c1=1\times 10^{-4}$, and $c2=9\times 10^{-4}$).  As a matter of fact, the calculation of mean value is a special kind of convolution, which is also named by average pooling, while variance operation actually involve twice operations of average pooling. It's obvious that the function of SSIM in Eq. (4-5) is differentiable, so the SSIM error can be efficiently back-propagated via optimization.

On the other hand, according to the Gamal and Cover theorem of \cite{m38, m37}, the MDGN network should pledge to have mutual information between two generated descriptions so that we can receive a acceptable reconstruction, even when only one description is got at the client. It's obvious that SSIM loss function keeps the two descriptions yielded by the MDGN network structurally similar. In the meantime, the two produced descriptions by neural networks are used as opposing labels to regularize the training of MDGN network. Consequently, the high-quality central reconstruction with two diverse descriptions can be guaranteed. Contrary to the SSIM loss, the distance loss function is utilized to keep the detail difference between two descriptions, which is written as:
\begin{equation}
\begin{split}
L_{dis} (f(\bm{I_A}),f(\bm{I_B}))=-\frac{1}{M \cdot N}\sum_{i}(||{\bm{I_A}}_i-{\bm{I_B}}_i||_L).
\end{split}
\end{equation}

For brevity latter, the content loss function and gradient difference loss function between two images $\bm{X}$ and $\bm{Y}$ are defined as:
\begin{equation}
\begin{split}
L_c(\bm{X},\bm{Y})=\frac{1}{M \cdot N}\sum_{i}(||\bm{X}_i-\bm{Y}_i||_L),
\end{split}
\end{equation}
\begin{equation}
\begin{split}
L_{gd}(\bm{X},\bm{Y})=\frac{1}{M \cdot N} \sum_{i} ((\sum_{s\in{\Omega_i}}||\nabla_s \bm{X}_i-\nabla_s \bm{Y}_i||_L)),
\end{split}
\end{equation}
where $\nabla_s$ is the $s$-th gradient between each pixel and $s$-th pixels among 8-neighbourhood $\Omega_i$. Here, L1-norm is chosen to produce sharper results than L2-norm, which has been reported in \cite{m42, m46}.

In the MDRN network, both content loss $L_c$ and gradient difference loss $L_{gd}$ supervise the learning of side reconstruction $R(c(\bm{I_A},\phi),\alpha_1)$, $R(c(\bm{I_B},\phi),\alpha_2)$ and central reconstruction $R(c(\bm{I_A},\phi),c(\bm{I_B},\phi),\alpha_3)$, which is presented as follows:
\begin{align}%\notag\\
&L_{MDRN}(\bm{I_A},\bm{I_B},\bm{I},\alpha)=\notag\\
&L_{c}(\bm{I},R(c(\bm{I_A},\phi),\alpha_1))+L_{gd}(\bm{I},R(c(\bm{I_A},\phi),\alpha_1))\notag\\
&+L_{c}(\bm{I},R(c(\bm{I_B},\phi),\alpha_2))+L_{gd}(\bm{I},R(c(\bm{I_B},\phi),\alpha_2))\notag\\
&+L_c(\bm{I},R(c(\bm{I_A},\phi),c(\bm{I_B},\phi),\alpha_3))\notag\\
&+L_{gd}(\bm{I},R(c(\bm{I_A},\phi),c(\bm{I_B},\phi),\alpha_3)).
\end{align}

In order to back-propagate the error from the MDRN network to the MDGN network, we learn MDVCN network to approximate the procedure from the lossless descriptions to the lossy description reconstruction. Both content loss and gradient difference loss are used to regularize the training of MDVCN network, which are given as follows:
\begin{align}%\notag\\
&L_{MDVCN}(\bm{I_A},\bm{I_B},\theta)=\notag\\
&L_{c}(\bm{\tilde{I}_A},V(\bm{I_A},\theta_1))+L_{gd}(\bm{\tilde{I}_A},V(\bm{I_A},\theta_1)) \notag\\
&+L_{c}(\bm{\tilde{I}_B},V(\bm{I_B},\theta_2))+L_{gd}(\bm{\tilde{I}_B},V(\bm{I_B},\theta_2))\notag\\
&+L_{c}(\bm{\tilde{I}},V(\bm{I_A},\bm{I_B},\theta_3))+L_{gd}(\bm{\tilde{I}},V(\bm{I_A},\bm{I_B},\theta_3)).
\end{align}

In addition to the aforementioned loss, we use MDVCN network to explicitly supervise the learning of the MDGN network or directly use gradient from MDVCN network as the gradient approximation from the standard codec. It's worth noticing that MDVCN network does not be used any more, once the whole training is finished, that is to say, only the MDRN network and the MDGN network during the testing are respectively leveraged to create multiple descriptions for compression and reconstruct these descriptions.
\subsection{Network architecture}
\begin{table}
\centering
    \caption{The structure of MDGN network}
    \label{tab:font-sizes}
    \scriptsize
% Table generated by Excel2LaTeX from sheet 'Sheet1'
\begin{tabular}{|c|c|c|c|c|c|}
\hline
                                         \multicolumn{ 6}{|c|}{MDGN Network} \\
\hline
{\bf Layer} &    {\bf k} &    {\bf s} & {\bf c-in} & {\bf c-out} & {\bf input} \\
\hline
    conv-1f &          9 &          1 &          1 &        128 &     $I$ \\
\hline
    conv-2f &          3 &          2 &        128 &        128 &     conv-1f \\
\hline
    conv-3f &          3 &          1 &        128 &        128 &     conv-2f \\
\hline
    conv-4f &          3 &          1 &        128 &        128 &     conv-3f \\
\hline
    conv-5A &          3 &          1 &        128 &        128 &     conv-4f \\
\hline
    conv-6A &          3 &          1 &        128 &        128 &     conv-5A \\
\hline
   convs-7A &          3 &          1 &        128 &        128 &     conv-6A \\
\hline
    conv-8A &          9 &          1 &        128 &          1 &     conv-7A \\
\hline
    conv-5B &          3 &          1 &        128 &        128 &     conv-4B \\
\hline
    conv-6B &          3 &          1 &        128 &        128 &     conv-5B \\
\hline
    conv-7B &          3 &          1 &        128 &        128 &     conv-6B \\
\hline
    conv-8B &          9 &          1 &        128 &          1 &     conv-7B \\
\hline
\end{tabular}

\end{table}
The MDGN network is composed with eight convolutional layers, which has one input stream, but two output streams, that is to say, the extracted feature maps with feature extraction network (FEN) from layer 1-4 are shared by generator network-A (GNA) and generator network-B (GNB). The FEN network has four convolutional layers, whose first layer's spatial kernel size is $9\times 9$ and other layers' is $3\times 3$. In the GNA and GNB networks, there are four convolutional layers with spatial kernel size $3\times 3$ except for the last layer with $9\times 9$. The large spatial kernel $9\times 9$ of convolutional layer in the first layer and last layer could further enlarge the receptive field of convolutional networks on the basis of small kernel $3\times 3$. Hence, image's context information is well considered during the generation of descriptions. The details about each layer in the MDGN network are listed in the Table I, where "k" represents the kernel size, "c-in" denotes the number of channel input, "c-out" is the total output map's number in the corresponding layer. Meanwhile, "conv" represents convolutional layer and "deconv" indicates the de-convolutional layer. From this table, it can be seen that all the layers employ stride step 1 except for the second convolutional layer with stride of 2. All the convolutional layers are activated by the ReLU activation function apart from the last layer in the MDGN network.
\begin{table}
\centering
    \caption{The structure of MDRN network}
    \label{tab:font-sizes}
    \scriptsize
% Table generated by Excel2LaTeX from sheet 'Sheet2'
\begin{tabular}{|c|c|c|c|c|c|}
\hline
                       \multicolumn{ 6}{|c|}{SRNA Network} \\
\hline
{\bf Layer} &    {\bf k} &    {\bf s} & {\bf c-in} & {\bf c-out} & {\bf input} \\
\hline
    conv-1a &          9 &          1 &          1 &        128 &      $\bm{\hat{I}_A}$ \\
\hline
    conv-2a &          3 &          1 &        128 &        128 &     conv-1a \\
\hline
    conv-3a &          3 &          1 &        128 &        128 &     conv-2a \\
\hline
    conv-4a &          3 &          1 &        128 &        128 &     conv-3a \\
\hline
    conv-5a &          3 &          1 &        128 &        128 &     conv-4a \\
\hline
    conv-6a &          3 &          1 &        128 &        128 &     conv-5a \\
\hline
    conv-7a &          3 &          1 &        128 &        128 &     conv-6a \\
\hline
  deconv-8a &          9 &          2 &        128 &          1 &     conv-7a \\
\hline
                       \multicolumn{ 6}{|c|}{SRNB Network} \\
\hline
{\bf Layer} &    {\bf k} &    {\bf s} & {\bf c-in} & {\bf c-out} & {\bf input} \\
\hline
    conv-1b &          9 &          1 &          1 &        128 &     $\bm{\hat{I}_B}$ \\
\hline
    conv-2b &          3 &          1 &        128 &        128 &     conv-1b \\
\hline
    conv-3b &          3 &          1 &        128 &        128 &     conv-2b \\
\hline
    conv-4b &          3 &          1 &        128 &        128 &     conv-3b \\
\hline
    conv-5b &          3 &          1 &        128 &        128 &     conv-4b \\
\hline
    conv-6b &          3 &          1 &        128 &        128 &     conv-5b \\
\hline
    conv-7b &          3 &          1 &        128 &        128 &     conv-6b \\
\hline
  deconv-8b &          9 &          2 &        128 &          1 &     conv-7b \\
\hline
                      \multicolumn{ 6}{|c|}{CRN Network} \\
\hline
{\bf Layer} &    {\bf k} &    {\bf s} & {\bf c-in} & {\bf c-out} & {\bf input} \\
\hline
    conv-1c &          9 &          1 &          2 &        128 &     $\bm{\hat{I}_A}$ and $\bm{\hat{I}_B}$  \\
\hline
    conv-2c &          3 &          1 &        128 &        128 &     conv-1c \\
\hline
    conv-3c &          3 &          1 &        128 &        128 &     conv-2c \\
\hline
    conv-4c &          3 &          1 &        128 &        128 &     conv-3c \\
\hline
    conv-5c &          3 &          1 &        128 &        128 &     conv-4c \\
\hline
    conv-6c &          3 &          1 &        128 &        128 &     conv-5c \\
\hline
    conv-7c &          3 &          1 &        128 &        128 &     conv-6c \\
\hline
  deconv-8c &          9 &          2 &        128 &          1 &     conv-7c \\
\hline
\end{tabular}

\end{table}
\begin{table}
\centering
    \caption{The structure of MDVCN network}
    \label{tab:font-sizes}
    \scriptsize
% Table generated by Excel2LaTeX from sheet 'Sheet2'
\begin{tabular}{|c|c|c|c|c|c|}
\hline
                       \multicolumn{ 6}{|c|}{VSRNA Network} \\
\hline
{\bf Layer} &    {\bf k} &    {\bf s} & {\bf c-in} & {\bf c-out} & {\bf input} \\
\hline
    conv-1a &          9 &          1 &          1 &        128 &      $\bm{I_A}$ \\
\hline
    conv-2a &          3 &          1 &        128 &        128 &     conv-1a \\
\hline
    conv-3a &          3 &          1 &        128 &        128 &     conv-2a \\
\hline
    conv-4a &          3 &          1 &        128 &        128 &     conv-3a \\
\hline
    conv-5a &          3 &          1 &        128 &        128 &     conv-4a \\
\hline
    conv-6a &          3 &          1 &        128 &        128 &     conv-5a \\
\hline
    conv-7a &          3 &          1 &        128 &        128 &     conv-6a \\
\hline
  deconv-8a &          9 &          2 &        128 &          1 &     conv-7a \\
\hline
                       \multicolumn{ 6}{|c|}{VSRNB Network} \\
\hline
{\bf Layer} &    {\bf k} &    {\bf s} & {\bf c-in} & {\bf c-out} & {\bf input} \\
\hline
    conv-1b &          9 &          1 &          1 &        128 &     $\bm{I_B}$ \\
\hline
    conv-2b &          3 &          1 &        128 &        128 &     conv-1b \\
\hline
    conv-3b &          3 &          1 &        128 &        128 &     conv-2b \\
\hline
    conv-4b &          3 &          1 &        128 &        128 &     conv-3b \\
\hline
    conv-5b &          3 &          1 &        128 &        128 &     conv-4b \\
\hline
    conv-6b &          3 &          1 &        128 &        128 &     conv-5b \\
\hline
    conv-7b &          3 &          1 &        128 &        128 &     conv-6b \\
\hline
  deconv-8b &          9 &          2 &        128 &          1 &     conv-7b \\
\hline
                      \multicolumn{ 6}{|c|}{VCRN Network} \\
\hline
{\bf Layer} &    {\bf k} &    {\bf s} & {\bf c-in} & {\bf c-out} & {\bf input} \\
\hline
    conv-1c &          9 &          1 &          2 &        128 &     $\bm{I_A}$ and $\bm{I_B}$ \\
\hline
    conv-2c &          3 &          1 &        128 &        128 &     conv-1c \\
\hline
    conv-3c &          3 &          1 &        128 &        128 &     conv-2c \\
\hline
    conv-4c &          3 &          1 &        128 &        128 &     conv-3c \\
\hline
    conv-5c &          3 &          1 &        128 &        128 &     conv-4c \\
\hline
    conv-6c &          3 &          1 &        128 &        128 &     conv-5c \\
\hline
    conv-7c &          3 &          1 &        128 &        128 &     conv-6c \\
\hline
  deconv-8c &          9 &          2 &        128 &          1 &     conv-7c \\
\hline
\end{tabular}

\end{table}

The MDRN network consists of SRNA network, SRNB network, and CRN network. In fact, we can let SRNA network and SRNB network share the same parameter set. Meanwhile, CRN network uses the outputs from the SRNA-network and SRNB-network to reconstruct the central images. But, in order to better back-propagate the errors from the MDRN network to the previous networks and avoid too deep networks for central reconstruction, we use three separate networks without cross connection and no weights sharing to respectively reconstruct side images and central image. They all use the eight convolutional layers. Seven convolutional layers and one deconvolution layer are used in the MDRN network so as to remove the coding artifacts and up-scale feature maps to the full-resolution at the same time. The obvious difference between them is that CRN network has two lossy descriptions as input while the two other networks only have one lossy descriptions as input. All the details are specified in the Table II, from which we can observe that the first and last convolutional layers use the $9\times 9$ spatial kernel to ensure the receptive field large enough, so that more spatial features are captured to better reconstruct the degraded descriptions. In addition, all the convolutional layers are activated by the ReLU, but the last layers of SRNA network, SRNB network, and CRN network are processed without any activation.

As described above, MDVCN network bridges the gap between MDGN network and MDRN network so that the errors of the reconstruction can be properly back-propagated from MDRN network to MDGN network. MDVCN network and MDRN network are designed to have same structure, because they can be seen as the same class of low-level image processing problems by learning. Thus, we have three virtual networks for MDVCN network: virtual side reconstruction network-A (VSRNA), virtual side reconstruction network-B (VSRNB), and virtual central reconstruction network (VCRN), whose network structures in the Table III are similar to the one's of MDRN network in the Table II. However, the inputs of MDVCN network and MDRN network are different, in which the former one takes the decoded lossy descriptions $\bm{\hat{I}_A}$ and $\bm{\hat{I}_B}$ as inputs, while the later one is fed with lossless multiple descriptions $\bm{I_A}$ and $\bm{I_B}$.

\subsection{Network learning}
Obviously, it's challenging to learn our whole framework directly, but our problem of learning multiple description neural networks can be separated into several sub-problems learning. In order to resolve these problems, we provide two learning ways for error back-propagation. These two ways are presented in the following and respectively referred to as \emph{learning algorithm-1} and \emph{learning algorithm-2}. Our \emph{learning algorithm-1} treats MDVCN network as feature function to build the reconstruction by fixing the parameter of MDVCN network so that reconstruction errors from MDVCN network can be back-propagated for the supervision of the MDGN network ahead of standard codec. It means that the MDGN network and the MDRN network are trained separately. On the contrary, our \emph{learning algorithm-2} uses MDVCN network's back-propagated error for MDGN network to approximately estimate the error from the codec without fixing any network's parameter, when explicitly training the MDGN network and the MDRN network simultaneously. The details about these two learning algorithms will be described next.

\subsubsection{Learning algorithm-1}

\begin{algorithm}[!h]
\caption{Learning Multiple Description Neural Networks}
\scriptsize
\begin{algorithmic}[1]
\renewcommand{\algorithmicrequire}{\textbf{Input:}}
\renewcommand{\algorithmicensure}{\textbf{Output:}}
\Require Ground truth image: $\bm{I}$; the number of iteration: $R$; the total number of images for training: $n$; the batch size during training: $m$;
\Ensure  The parameter sets of MDGN network and MDRN network: $\omega$, $\alpha$;
\State Initialize to produce multiple descriptions $\bm{I_A}$ and $\bm{I_B}$ by down-sampling for preparation of the training of MDRN network;
\State Initialize parameter sets: $\omega$, $\alpha$, $\phi$, $\theta$;
\For{$r=1$ to $ R$}
    \State Compress multiple descriptions $\bm{I_A}$ and $\bm{I_B}$ by standard codec with $\phi$;
    \For{$epoch=1$ to $p$}
        \For{$i=1$ to floor$(n/m)$}
            \State Update the parameter set of $\alpha$ to train the MDRN network according to
            \State the minimization of the Eq. (9) with $i$-th batch images;
        \EndFor
    \EndFor
    \State Generate the multiple descriptions reconstruction dataset $\bm{\tilde{I}_A}$ and $\bm{\tilde{I}_B}$ with the
    \State parameter set of $\alpha$ by the MDGN network;
    \For{$epoch=1$ to $p$}
        \For{$j=1$ to floor$(n/m)$}
            \State Update the parameter set of $\theta$ by training MDVCN network to minimize
            \State the Eq. (10) with $j$-th batch images from $\bm{\tilde{I}_A}$ and $\bm{\tilde{I}_B}$ dataset;
        \EndFor
    \EndFor
    \For{$epoch=1$ to $q$}
        \For{$l=1$ to floor$(n/m)$}
            \State Update the parameter set of $\omega$ with fixed $\theta$ to train the MDGN network
            \State based on minimization the Eq. (3) and Eq. (9) with $l$-th batch images;
        \EndFor
    \EndFor
    \State Generate the multiple descriptions images $\bm{I_A}$ and $\bm{I_B}$ with the parameter set
    \State of $\omega$ by the MDGN network;
\EndFor
\For{$epoch=1$ to $p$}
    \For{$i=1$ to floor$(n/m)$}
        \State Update the parameter set of $\alpha$ by training the MDRN network to minimize
        \State Eq. (9) with $i$-th batch images;
    \EndFor
\EndFor
\State \textbf{return} $\omega$, $\alpha$;
\end{algorithmic}
\end{algorithm}

To back-propagate the error from the MDRN network to MDGN network, we decompose the learning problem of MDGN network, MDRN network and MDVCN network once in Eq. (1) into three separate subproblem learning, but they depends on each other closely. Specifically, we first initialize all the parameter sets mentioned previously, and multiple descriptions $\bm{I_A}$ and $\bm{I_B}$ dataset by down-sampling for the training of MDRN network and compress this dataset. Secondly, the parameter set of $\alpha$ is updated by training MDRN network based on minimization of the Eq. (9). Then, we generate multiple descriptions reconstruction images $\bm{\tilde{I}_A}$, $\bm{\tilde{I}_B}$, and $\bm{\tilde{I}}$ dataset with the parameter set of $\alpha$ of MDRN network. This reconstruction dataset $\bm{\tilde{I}_A}$, $\bm{\tilde{I}_B}$, and $\bm{\tilde{I}}$ can be used to train MDVCN network by updating the parameter set of $\theta$ based on the minimization of the Eq. (10). Next, we update the parameter set of $\omega$ with fixed $\theta$ to train MDGN network according to the minimization of the Eq. (3) and Eq. (9). After training MDGN network, the multiple descriptions images $\bm{I_A}$ and $\bm{I_B}$ are generated with the parameter set $\omega$ of MDGN network and then start the next iteration. The details about \emph{learning algorithm-1} are summarized in the \textbf{Algorithm-1}.

\subsubsection{Learning algorithm-2}

Different from \emph{our learning algorithm-1}, we separate the whole framework learning as two sub-problem learning: the sub-problem of simultaneously learning MDGN network and MDRN network, and the learning sub-problem of MDVCN network. Concretely, the parameter sets of MDGN network and MDRN network: $\omega$, $\alpha$ are learned by the optimization with gradient descent method at the same time. After feeding input data into MDGN network to produce multiple descriptions $\bm{I_A}$ and $\bm{I_B}$ and compressing them with standard codec, MDRN network are used to reconstruct these compressed multiple descriptions $\bm{\hat{I}_A}$ and $\bm{\hat{I}_B}$. Meanwhile, the lossless multiple descriptions $\bm{I_A}$ and $\bm{I_B}$ are fed into MDVCN network. This is feed-forward propagation of our deep convolution neural networks, but the error from the MDRN network is blocked by the codec. Here, we can explicitly use the error from MDVCN network as the approximate error from the codec. Thus, we can simultaneously update MDGN network and MDRN network. The whole process is detailed in the \textbf{Algorithm-2}.
\begin{algorithm}[!h]
\caption{Learning Multiple Description Neural Networks}
\scriptsize
\begin{algorithmic}[1]
\renewcommand{\algorithmicrequire}{\textbf{Input:}}
\renewcommand{\algorithmicensure}{\textbf{Output:}}
\Require Ground truth image: $\bm{I}$; the number of iteration: $T$; the total number of images for training: $n$; the batch size during training: $m$;
\Ensure  The parameter sets of MDGN network and MDRN network: $\omega$, $\alpha$;
\State Initialize parameter sets: $\omega$, $\alpha$, $\phi$, $\theta$;
\State Pre-train MDVCN network;
\For{$t=1$ to $T$}
    \For{$epoch=1$ to $l$}
        \For{$i=1$ to floor$(n/m)$}
            \State $a):$ Generate multiple descriptions $\bm{I_A}$ and $\bm{I_B}$ with the the parameter
            \State set of $\omega$; Then, compress multiple descriptions $\bm{I_A}$ and $\bm{I_B}$ with standard
            \State codec with $\phi$;
            \State $b):$ Update the parameter set of $\omega$ and $\alpha$ by training the MDGN network
            \State and MDRN network simultaneously by minimizing Eq. (3) and  Eq. (9)
            \State with $i$-th batch images. Note that the MDGN network uses errors from
            \State MDVCN network for back-propagation to update parameter set of $\omega$;
            \State $c):$ Generate multiple descriptions reconstruction images $\bm{\tilde{I}_A}$ and $\bm{\tilde{I}_B}$
            \State with the parameter set of $\alpha$ by MDRN network;
            \State $d):$ Update the parameter set of $\theta$ by training MDVCN network based
            \State on minimization of Eq. (10) with $i$-th batch images with $\bm{\tilde{I}_A}$ and $\bm{\tilde{I}_B}$;
        \EndFor
    \EndFor
\EndFor
\State \textbf{return} $\omega$, $\alpha$;
\end{algorithmic}
\end{algorithm}

After comparing \emph{learning algorithm-1} with \emph{learning algorithm-2}, we can see that the training stability of the second one relies on whether pre-trained MDVCN network is well trained or not. Meanwhile, this network also has great impacts on the learning of MDGN network, because the bad accuracy of approximated error propagation from MDVCN network will results in the insufficiency of multiple description generation. On the contrary, the first algorithm is more easily implemented in any neural network platform, because there is no any changes in the process of neural network's optimization. Meanwhile, the performance of learning algorithm-1 tends to be more stable than the second one due to the reliable dependency among three neural networks. It comes from a fact that the good training of MDRN network will directly lead to the good training of MDVCN network, and then MDVCN network will give a supervision of the MDGN network. Conclusively, both of them can resolve the learning problem of multiple description neural networks, but the learning algorithm-1 is more practical, so we use it to illustrate the efficiency of the whole framework in the experimental sections.

\begin{figure}[!ht]
\centering
\includegraphics[width=3.5in]{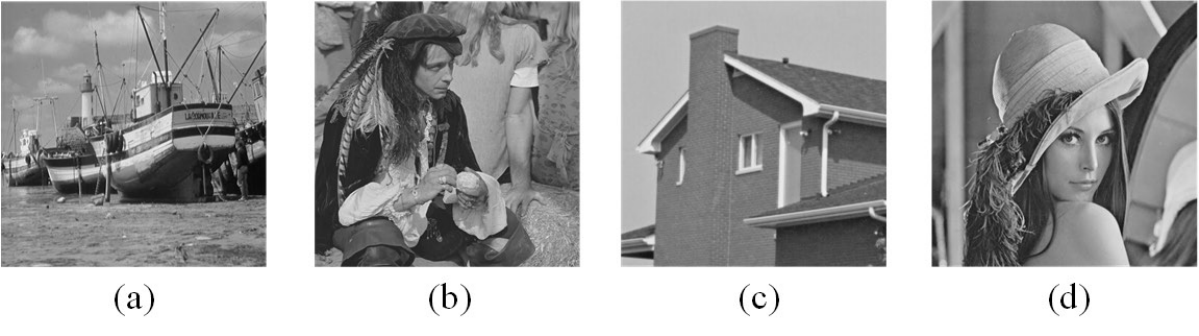}
\caption{The data-set is used for our testing}
\label{Fig2}
\end{figure}
\section{Experimental results}
We evaluate the proposed method against eight baselines with state-of-the-art artifacts removal techniques \cite{m21,m22,m23,m24} and advanced super-resolution based on very deep convolutional neural networks, such as \cite{m25,m45}. Note that there are 20 convolutional layers used for super-resolution in \cite{m25,m45}. Four baselines "MDB1a-MDB4a" are formed with the techniques of artifacts removal \cite{m21,m22,m23,m24} and very deep convolutional neural network based super-resolution \cite{m25}. Meanwhile, super-resolution of \cite{m45} are combined with artifacts removal \cite{m21,m22,m23,m24} to build four other baselines "MDB1b-MDB4b".  Furthermore, in order to fully demonstrate the efficiency of the proposed method, we form a baseline model, which is denoted as "Our-base", when replacing MDGN network to generate multiple descriptions with the poly-phase down-sampling technique in \cite{m7}. For simplicity, the proposed method is marked as "Ours". Besides, the training of the proposed framework will be in detail described next.
\begin{figure*}[!ht]
\centering
\includegraphics[width=7in]{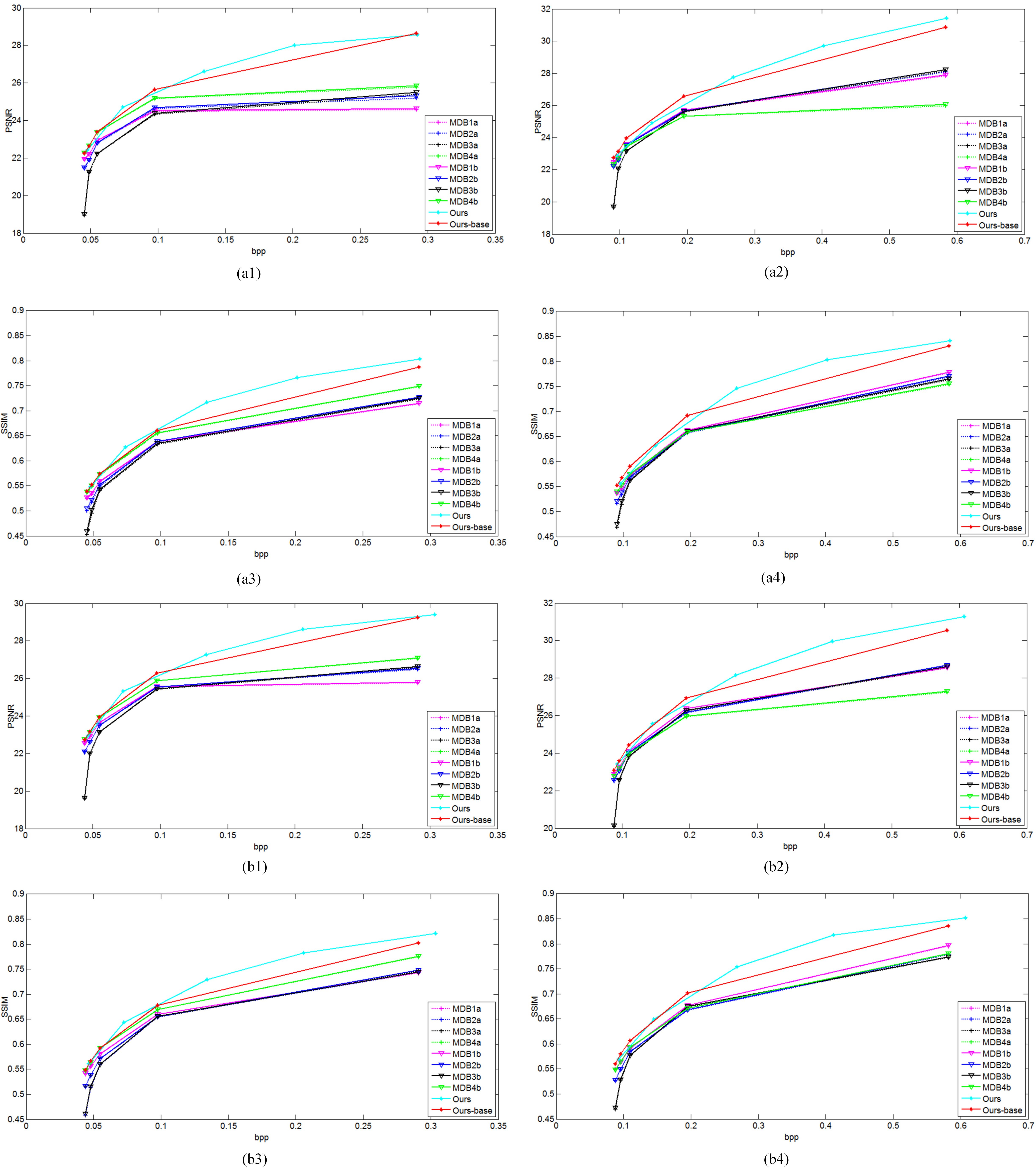}
\caption{The side reconstruction and central reconstruction objective measurement comparison on PSNR and SSIM for several state-of-the-art approaches. (a1,b1) are respectively the side-reconstruction PSNR results of image (a) and (b) in Fig. \ref{Fig2}, (a2,b2) are the central reconstruction PSNR results of image (a) and (b) in Fig. \ref{Fig2}, (a3,b3) are respectively the side reconstruction SSIM results of image (a) and (b) in Fig. \ref{Fig2}, (a4,b4) are the central reconstruction SSIM results of image (a) and (b) in Fig. \ref{Fig2}}
\label{Fig3}
\end{figure*}

\begin{figure*}[!ht]
\centering
\includegraphics[width=7in]{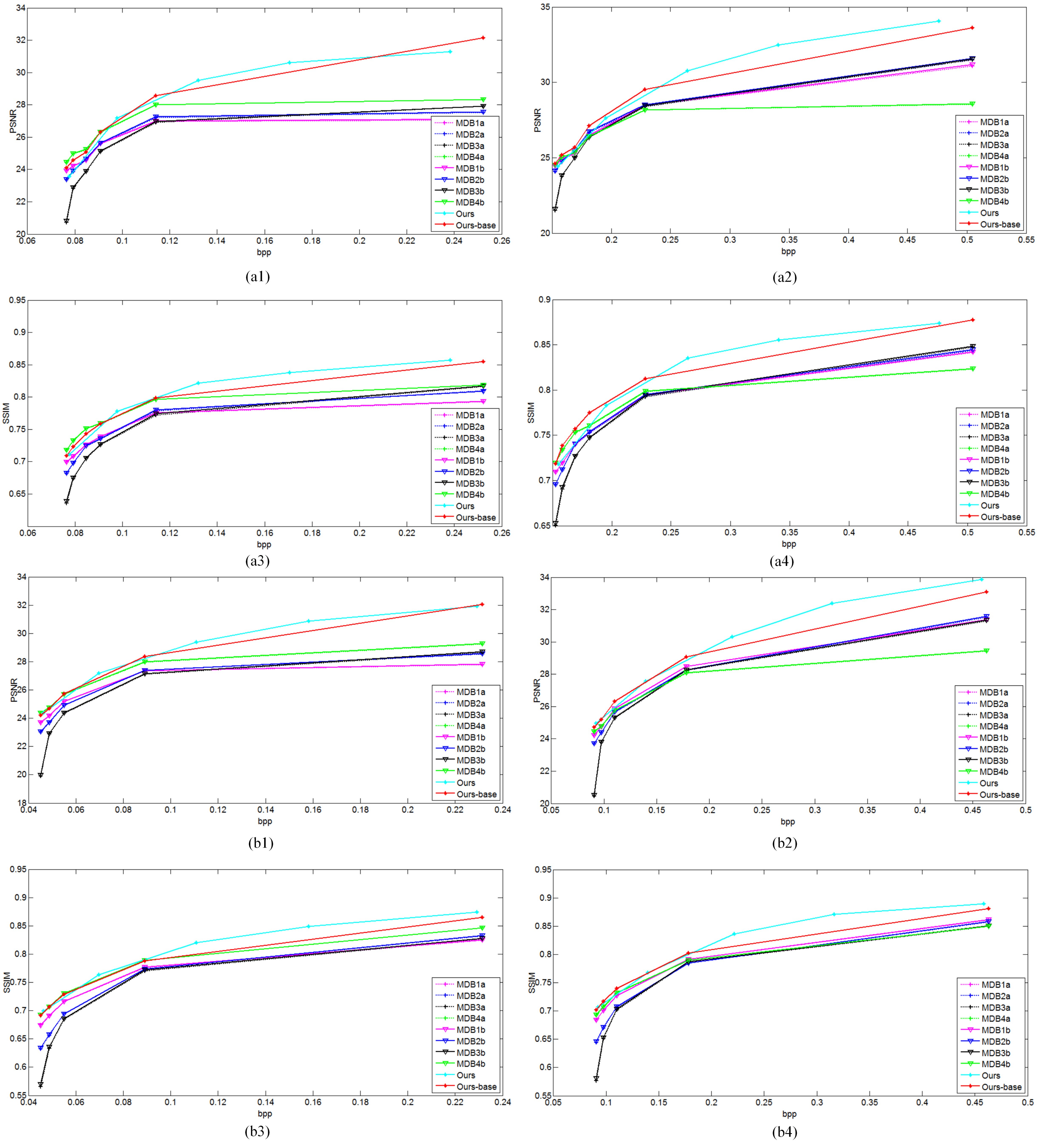}
\caption{The side-reconstruction and central reconstruction objective measurement comparison on PSNR and SSIM for several state-of-the-art approaches. (a1,b1) are respectively the side-reconstruction PSNR results of image (c) and (d) in Fig. \ref{Fig2}, (a2,b2) are the central reconstruction PSNR results of image (c) and (d) in Fig. \ref{Fig2}, (a3,b3) are respectively the side-reconstruction SSIM results of image (c) and (d) in Fig. \ref{Fig2}, (a4,b4) are the central reconstruction SSIM results of image (c) and (d) in Fig. \ref{Fig2}}
\label{Fig4}
\end{figure*}

\begin{figure*}[!ht]
\centering
\includegraphics[width=7in]{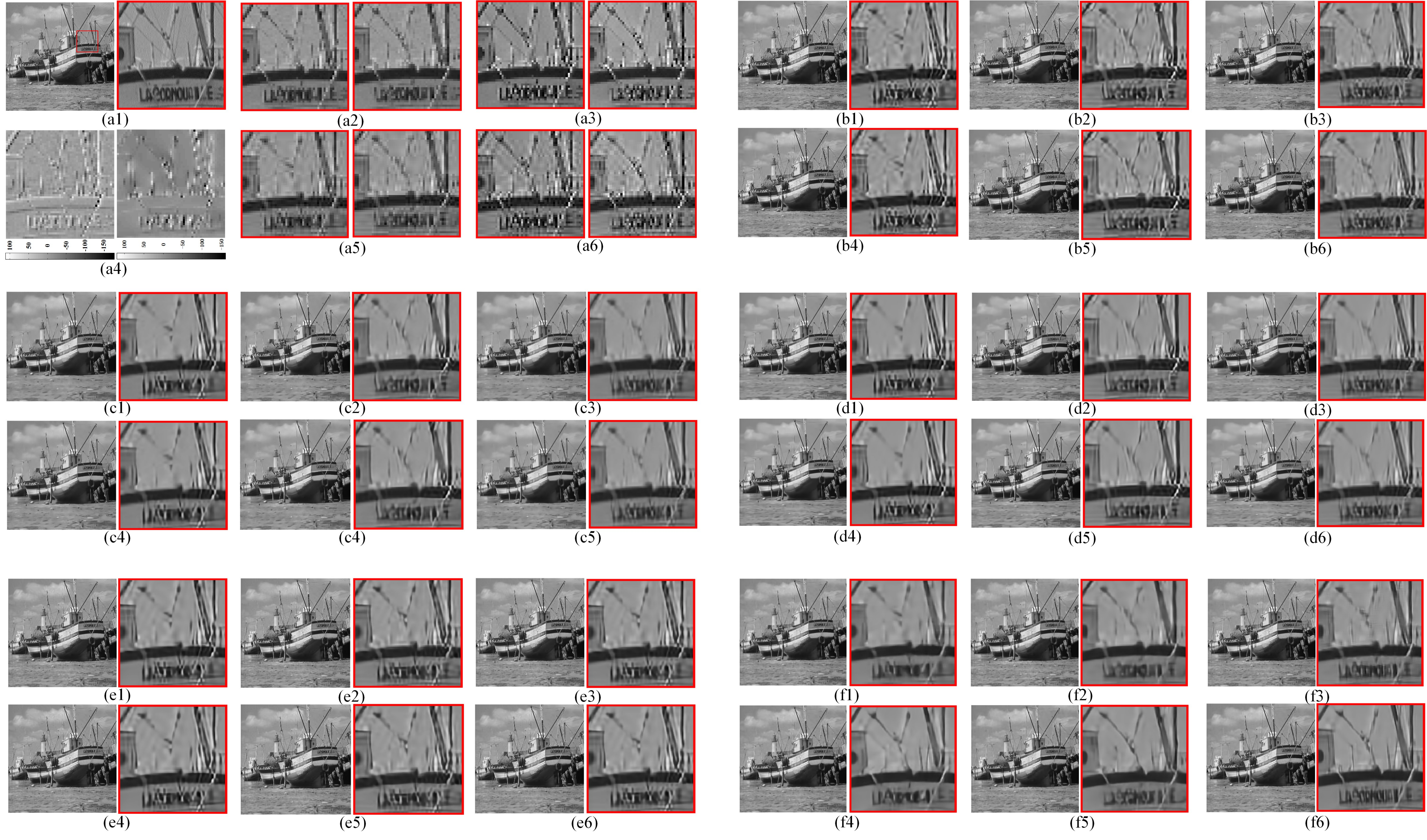}
\caption{The visual comparison of different methods for (a) in Fig. \ref{Fig2}. (a1) input image (left) and the enlargement in the red line boxed region of input image (right), (a2) multiple description created by a poly-phase down-sampling technique, (a3) multiple description generated by the proposed MDGN network, (a4) left image is the difference between a pair of image in (a2) and the right image is difference between a pair of image in (a3), (a5) is the compressed image in (a2), (a6) is the compressed image in (a3); (b-f) description reconstruction images, where the (b1-f1, b2-f2) and (b4-f4, b5-f5) are the side reconstruction images, the (b3-f3) and (b6-f6) are the central reconstruction images; (b1-b3) MDB1a(24.574/0.714/0.292(s) and 27.849/0.778/0.583(c)), (b4-b6) MDB1b(24.637/0.715/0.292(s) and 27.849/0.778/0.583(c)), (c1-c3) MDB2a(25.202/0.725/0.292(s) and 28.088/0.769/0.583(c)), (c4-c6) MDB2b(25.318/0.727/0.292(s) and 28.215/0.771/0.583(c)), (d1-d3) MDB3a(25.399/0.723/0.292(s) and 28.129/0.763/0.583(c)), (d4-d6) MDB3b(25.507/0.726/0.292(s) and 28.231/0.765/0.583(c)); (e1-e3) MDB4a(25.785/0.748/0.292(s) and 25.969/0.754/0.583(c)), (e4-e6) MDB4a(25.847/0.749/0.292(s) and 26.051/0.756/0.583(c));(f1-f3) Ours-base(28.641/0.787/0.292(s) and 30.870/0.831/0.583(c)), (f4-f6) Ours(28.577/0.803/0.292(s) and 31.410/0.842/0.585(c)). (Note that the red line boxed regions in (b-f) represent the part regions enlarged from the corresponding full resolution images like (a1); the real image size of (a2-a6) is half of input image's size, while all the other images have the same size as the input image)}
\label{Fig5}
\end{figure*}

\begin{figure*}[!ht]
\centering
\includegraphics[width=7in]{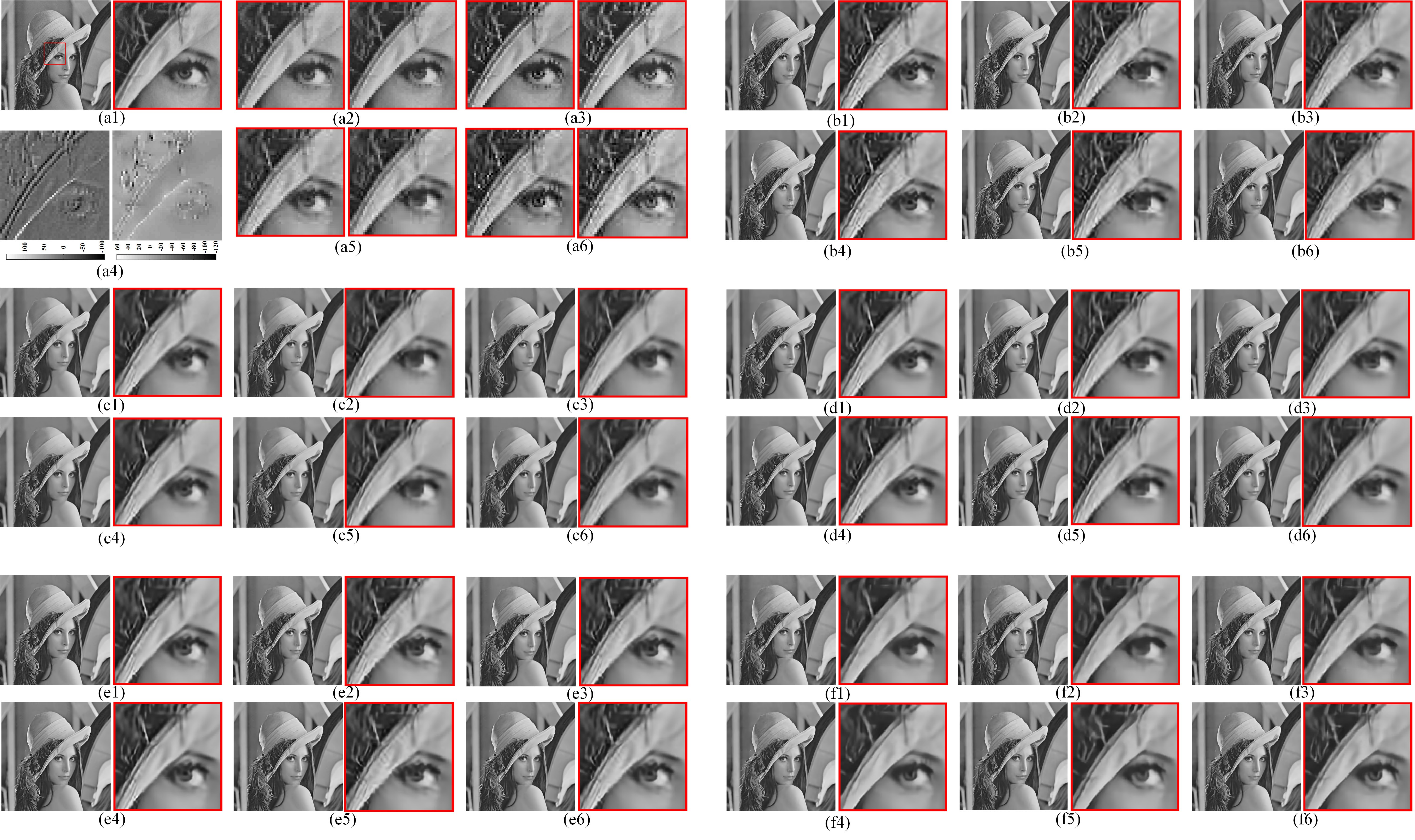}
\caption{The visual comparison of different methods for (d) in Fig. \ref{Fig2}. (a1) input image (left) and the enlargement in the red line boxed region of input image (right), (a2) multiple description created by a poly-phase down-sampling technique, (a3) multiple description generated by the proposed MDGN network, (a4) left image is the difference between a pair of image in (a2) and the right image is difference between a pair of image in (a3), (a5) is the compressed image in (a2), (a6) is the compressed image in (a3); (b-f) description reconstruction images, where the (b1-f1, b2-f2) and (b4-f4, b5-f5) are the side reconstruction images, the (b3-f3) and (b6-f6) are the central reconstruction images; (b1-b3) MDB1a(27.808/0.824/0.232(s) and 31.319/0.861/0.463(c)), (b4-b6) MDB1b (27.843/0.825/0.232(s) and 31.367/0.862/0.463(c)), (c1-c3) MDB2a(28.524/0.832/0.232(s) and 31.526/0.857/0.463(c)), (c4-c6) MDB2b(28.579/0.833/0.232(s) and 31.601/0.858/0.463(c)), (d1-d3) MDB3a(28.642/0.827/0.232(s) and 31.288/0.850/0.463(c)), (d4-d6) MDB3b(28.700/0.828/0.232(s) and 31.352/0.851/0.463(c)); (e1-e3) MDB4a(29.244/0.846/0.232(s) and 29.428/0.850/0.463(c)), (e4-e6) MDB4a(29.270/0.846/0.232(s) and 29.471/0.850/0.463(c)); (f1-f3) Ours-base(32.040/0.865/0.232(s) and 33.098/0.881/0.463(c)), (f4-f6) Ours(31.913/0.874/0.229(s) and 33.865/0.889/0.458(c))). (Note that the red line boxed regions in (b-f) represent the part regions enlarged from the corresponding full resolution images like (a1); the real image size of (a2-a6) is half of input image's size, while all the other images have the same size as the input image)}
\label{Fig6}
\end{figure*}

\subsection{Training data and implementation details}
Our whole framework is implemented in the platform of TensorFlow \cite{m39} with \textbf{Algorithm-1}. The 400 images with size 180x180 from \cite{m40} are used as our training data-set, which are augmented by cropping, flipping, and rotating image to build our training data set. There are the total number 3200 of image patches with size of 160x160 used for our framework's training. Four images in Fig. 2 are used to evaluate the efficiency of the proposed method for testing. Our framework is trained with the Adam optimization method \cite{m47}. The parameters for Adam optimization are set to be $\beta_1=0.9$, $\beta_2=0.999$. The learning rate of training is initially set as 0.0001, but the learning rate decays to be half of the initial one when the training step reaches 3/5 of total step. Once the training step reaches 4/5 of total step, it reduces to be 1/4 of the initial one. The multiple descriptions are compressed by standard JPEG codec with $QF$ to be 2, 6, 10, 20, and 40 for the proposed framework during the training and testing. The multiple descriptions for ”MDB1a-MDB4a” and ”MDB1b-MDB4b” as well as "Our-base" are compressed with the $QF$ set 2, 3, 4, 10, and 50.

\subsection{Comparisons with several baselines}
To validate the efficiency of the proposed framework, we employ the Peak Signal to Noise Ratio (PSNR) and SSIM to measure the objective quality. The multiple description artifacts removal results with Foi's \cite{m21}, BM3D \cite{m24}, DicTV \cite{m22} and CONCOLOR \cite{m23} are got with strict usage of the author's open codes according to the parameter settings in their papers. Meanwhile, for image super-resolution in \cite{m25, m45}, we use their official provided model to enlarge these multiple description after artifacts removal so as to guarantee the advances of eight baselines, when comparing with the proposed method.

From the comparison in Fig. \ref{Fig3} and Fig. \ref{Fig4}, it can be seen that ours-base has better performance on SSIM for the side reconstruction and cental reconstruction against eight baselines MDB1a-MDB4a and MDB1b-MDB4b in the full range. In the most cases, the PSNR measurement of ours-base is better than eight baselines MDB1a-MDB4a and MDB1b-MDB4b. Only at the very low bit-rate, the PSNR of ours-base has slight smaller than MDB4a and MDB4b, but our-base with higher SSIM measurement has priority than MDB4a and MDB4b. This comes from that the structural preservation of image is more significant than detail preservation at the very low bit-rate, as shown in Fig. \ref{Fig3} and Fig. \ref{Fig4}.

Compared to the ours-base, the proposed method has more PSNR and SSIM gains in the most cases, especially at the high bit-rate. Because the proposed method in this paper focuses on the appearance similarity but details difference for multiple descriptions generation without apparent structural distance loss to regularize the training at the very low bit-rate, the proposed method has a litter lower PSNR gains than ours-base in some cases. For the improvement of the proposed method at low bit-rate, the first way is to replace direct description distance loss with structural distance loss during training. Another feasible way is to employ 4x resolution reduction when generating the descriptions with MDGN network and compressing descriptions at the very low bit-rate, but larger $QF$ is used for the proposed method like our previous work \cite{m30}.

Among these baselines, MDB4-a and MDB4-b defeat against MDB1a-MDB3a and MDB1b-MDB3b on PSNR and SSIM measurement, when comparing side description reconstruction quality. But for the central reconstruction, MDB4a and MDB4b can not compete with the MDB1a-MDB3a and MDB1b-MDB3b. MDB3a and MDB3b have the best PSNR performance of the central reconstruction among the eight baselines. MDB1a-MDB3a have very similar performance on central reconstruction. Although the literature of \cite{m45} has reported that their approach has greater PSNR gains than \cite{m25} for general image super-resolution, the performance of \cite{m45} is slight better than the one's of \cite{m25}, when these super-resolution approaches are used for description's resolution enhancement after artifacts removal, which can be found in Fig. \ref{Fig3} and Fig. \ref{Fig4}, when comparing MDB1a-MDB4a with MDB1b-MDB4b.

We have compared the visual quality of the proposed method with different methods' for multiple description coding based on deep convolutional neural networks, which is displayed in Fig. \ref{Fig5} and Fig. \ref{Fig6}. In these figures, MDB1a(24.574/0.714/0.292(s) and 27.849/0.778/0.583(c)) represents the measurements of PSNR/SSIM/bpp for side reconstruction and central reconstruction based on the approach of MDB1a. Similarly, other methods can be denoted in this way. Our MDGN network-produced descriptions, as displayed in Fig. \ref{Fig5}-(a3) and Fig. \ref{Fig6}-(a3), maintain more important details than the ones generated with the poly-phase down-sampling technique \cite{m7}, even after image compression. The differences between these pairs of descriptions are exhibited in Fig. \ref{Fig5}-(a4) and Fig. \ref{Fig6}-(a4), from which it can be observed that the proposed method tends to keep the description distance on the details and has less structural difference preservation. Furthermore, the descriptions from our MDGN network tend to highlight obvious feature pixels for all the descriptions in order to protect the key features. Therefore, the protected feature of lossy descriptions always can be kept, although they are possibly badly smoothed and contaminated by compression, as shown in Fig. \ref{Fig5}-(a5-a6) and Fig. \ref{Fig6}-(a5-a6).

The side reconstruction images and cental reconstruction images have been displayed in Fig. \ref{Fig5}-(b-f) and Fig. \ref{Fig6}-(b-f). From these figures, it can be clearly seen that the side reconstruction images and cental reconstruction images with the proposed method look more natural and have more detail preservation than eight baselines MDB1a-MDB4a, MDB1b-MDB4b, and our-base. Our-base has better performance than the eight baselines. Among these baselines, both MDB4a and MDB4b keep more details than MDB1a-MDB3a, MDB1b-MDB3b, which can be seen in  Fig. \ref{Fig5}-(b-e) and Fig. \ref{Fig6}-(b-e). From the above objective and visual comparisons, it can be concluded that it's very important to emphasize on significant context features when automatically generating appearance-similar but details-different descriptions with convolutional neural networks, as compared to the poly-phase down-sampling technique. Meanwhile, the better descriptions always benefit the better side and central description reconstruction.

\section{Conclusion}
In this paper, we introduce multiple description image coding based on deep convolutional neural networks. First, multiple description network is employed to automatically yield valid multiple descriptions. Then, these multiple descriptions are compressed by standard codec so that our whole framework is compatible with standard codec. Thirdly, we use multiple description reconstruction network to enhance these descriptions and restore them to be full resolution for the reconstruction of the compressed multiple descriptions. Besides, two learning algorithms are provided to train our whole framework. Moreover, both distance loss and SSIM loss are combined together to train the multiple description generator networks in order to make sure that the generated multiple descriptions are diverse, but they have shared structures information.
% Can use something like this to put references on a page
% by themselves when using endfloat and the captionsoff option.
%\ifCLASSOPTIONcaptionsoff
%  \newpage
%\fi

% trigger a \newpage just before the given reference
% number - used to balance the columns on the last page
% adjust value as needed - may need to be readjusted if
% the document is modified later
%\IEEEtriggeratref{8}
% The "triggered" command can be changed if desired:
%\IEEEtriggercmd{\enlargethispage{-5in}}

% references section

% can use a bibliography generated by BibTeX as a .bbl file
% BibTeX documentation can be easily obtained at:
% http://mirror.ctan.org/biblio/bibtex/contrib/doc/
% The IEEEtran BibTeX style support page is at:
% http://www.michaelshell.org/tex/ieeetran/bibtex/
\bibliographystyle{IEEEtran}
\bibliography{IEEEfull,MDCNN}
\end{document}